\documentclass[12pt,tightenlines]{revtex4}

\usepackage{graphicx}
\usepackage{color} 
\usepackage{amsmath}
\usepackage{amssymb}

\def\Ket#1{\left|#1\right\rangle} 
\def\Bra#1{\left\langle#1\right|}
\def\KetBra#1#2{\Ket{#1}\!\Bra{#2}} 
\def\Proj#1{\KetBra{#1}{#1}}
\def\ProjInd#1#2{\Ket{#1}_{\!#2}\!\Bra{#1}}  
\def\Eins{\mathbf{1}} 

\def\ie{i.\,e.\ }

\def\bell#1{\ensuremath{\mathcal{B}_{#1}}}
\def\rhospace{\ensuremath{\mathcal{B}(\mathcal{H})}}

\newlength{\figurewidth}

\def\botrule{\hline}
\def\toprule{\hline}
\def\colrule{\hline}

\begin{document}

\title{Entanglement purification with noisy apparatus can be used
  to factor out an eavesdropper}


\author{Hans Aschauer}
\email{Hans.Aschauer@Physik.uni-muenchen.de}
\author{Hans J. Briegel}
\email{Hans.Briegel@Physik.uni-muenchen.de}

\affiliation{Sektion Physik, Ludwig-Maximilians-Universit\"at,
  Theresienstr.\ 37, D-80333 M\"unchen, Germany}

\date{\today}

\begin{abstract}
  We give a proof that entanglement purification, even with
  noisy apparatus, is sufficient to disentangle an eavesdropper (Eve)
  from the communication channel. Our proof applies to all possible
  attacks (individual and coherent). Due to the quantum nature of the
  entanglement purification protocol, it is also possible to use the
  obtained quantum channel for secure transmission of quantum
  information.  
\end{abstract}
\pacs{PACS: 3.67.Dd, 3.67.Hk, 3.65.Bz} 
\maketitle

\maketitle


\setlength{\figurewidth}{0.7\linewidth}

\section{Introduction}
\label{sec:Intro}
Quantum communication exploits the quantum properties of its
information carriers for communication purposes such as the
distribution of secure cryptographic keys in quantum cryptography
\cite{bb84,ekert91} and the communication between distant quantum
computers in a network \cite{cirac98}. A central problem of quantum
communication is how to faithfully transmit unknown quantum states
through a noisy quantum channel \cite{schumacher_noisy_channel}. While
information is sent through such a channel (for example an optical
fiber), the carriers of the information interact with the channel,
which gives rise to the phenomenon of decoherence and absorbtion; an
initially pure quantum state becomes a mixed state when it leaves the
channel. For quantum communication purposes, it is however necessary
that the transmitted qubits retain their genuine quantum properties,
for example in form of an entanglement with qubits on the other side
of the channel.

In quantum cryptography, noise in the communication channel plays a
crucial role: In the worst-case scenario, all noise in the channel
is attributed to an eavesdropper, who manipulates the qubits in order to
gain as much information on their state as possible, while introducing
only a moderate level of noise. 

To deal with this situation, two different techniques have been
developed: \emph{Classical privacy amplification} allows the
eavesdropper to have partial knowledge about the raw key built up
between the communicating parties Alice and Bob. From the raw key, a
shorter key is ``distilled'' about which Eve has vanishing (\ie
exponentially small in some chosen security parameter)
knowledge. Despite of the simple idea, proofs taking into account all
eavesdropping attacks allowed by the laws of quantum mechanics have
shown to be technically involved \cite{mayers,biham,hitoshi}.
Recently, Shor and Preskill \cite{shor} have given a simpler physical
proof relating the ideas in \cite{mayers,biham} to quantum error
correcting codes \cite{CSS} and, equivalently, to one-way entanglement
purification protocols. \emph{Quantum privacy amplification} (QPA)
\cite{deutsch96}, on the other hand, employs a two-way entanglement
purification recurrence protocol \cite{bennett96a} that eliminates any
entanglement with an eavesdropper by creating a few perfect EPR pairs
out of many imperfect (or impure) EPR pairs. The perfect EPR pairs can
then be used for secure key distribution in entanglement-based quantum
cryptography \cite{deutsch96,ekert91,mermin1992}. In principle, this
method guarantees security against any eavesdropping attack. However,
the problem is that the QPA protocol assumes ideal quantum operations.
In reality, these operations are themselves subject to noise. As shown
in \cite{briegel,duer_briegel,giedke}, there is an upper bound
$F_{\text{max}}$ for the achievable fidelity of EPR pairs which can be
distilled using noisy apparatus. \emph{A priori}, there is no way to
be sure that there is no residual entanglement with an eavesdropper.
This problem could be solved if Alice and Bob had fault tolerant
quantum computers at their disposal, which could then be used to
reduce the noise of the apparatus to any desired level. This was an
essential assumption in the security proof given by Lo and Chau
\cite{lo}.

In this paper, we show that the standard two-way entanglement
purification protocols alone, which have been developed by Bennett
\emph{at al.} \cite{bennett96,bennett96a} and later by Deutsch \emph{et
  al.} \cite{deutsch96},  with some minor modifications to
accomodate certain security aspects as discussed below, can
be used to efficiently establish a \emph{perfectly private quantum
channel}, even when both the physical channel connecting the parties
and the local apparatus used by Alice and Bob are noisy.

In Section \ref{sec:QRandQPA} we will briefly review the concepts of
entanglement purification. Section \ref{sec:fact_of_eve} will give the
main result of our work: we prove that it is possible to \emph{factor
out} an eavesdropper using EPP, even when the apparatus used by Alice
and Bob is noisy.  We conclude the paper with a
discussion in Section \ref{sec:dicussion}.

\section{Entanglement purification}
\label{sec:QRandQPA}

\label{sec:ent_purification}
\begin{figure}[htbp]
  \begin{center}
    \includegraphics{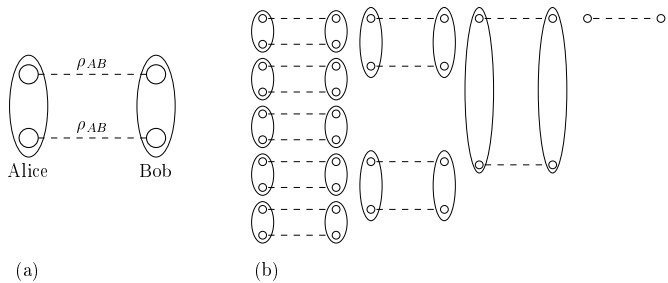}
    \caption{    
      \label{fig:ep_prot_and_proc}
      The entanglement purification protocol (a) and the
      entanglement distillation process (b).} 
  \end{center}
\end{figure}

As two-way entanglement purification protocols (2--EPP) play an
important role in this paper, we will briefly review one example of a
a recurrence protocol which was described in \cite{deutsch96}, and
called \emph{quantum privacy amplification} (QPA) by the authors. It
is important to note that we distinguish the entanglement purification
\emph{protocol} from the distillation \emph{process}: the first
consists of probabilistic local operations (unitary rotations and
measurements), where two pairs of qubits are combined, and either one
or zero pairs are kept, depending on the measurement outcomes. The
latter, on the other hand, is the procedure where the purification
protocol is applied to large ensemble of pairs recursively (see
Fig.~(\ref{fig:ep_prot_and_proc})).

In the quantum privacy amplification 2--EPP, two pairs of qubits,
shared by Alice and Bob, are considered to be in the state
\(\rho_{A_1B_1} \otimes \rho_{A_2B_2}\). Without loss of generality
(see later), we may assume that the state of the pairs is of the
Bell-diagonal form,
\begin{equation}
  \label{eq:bell_diagonal}
  \rho_{AB} = A \Proj{\Phi^+} + B \Proj{\Psi^-} + C \Proj{\Psi^+} + D
  \Proj{\Phi^-}.
\end{equation}
Following \cite{deutsch96}, the protocol consists of three steps: 
\begin{enumerate}
\item Alice applies to her qubits a \(\pi/2\) rotation, $U_x$, Bob a
  \(-\pi/2\) rotation about the \(x\) axis, $U_x^{-1}$.
\item Alice and Bob perform the bi-lateral CNOT operation
  \[\mathrm{BCNOT_{A_1B_1}^{A_2B_2}} = \mathrm{CNOT}_{A_1}^{A_2}
  \otimes  \mathrm{CNOT}_{B_1}^{B_2} \]
  on the four qubits.
\item Alice and Bob measure both qubits of the target pair \(A_2B_2\)
  of the BCNOT operation in the $z$ direction. If the measurement
  results coincide, the source pair \(A_1B_1\) is kept, otherwise it
  is discarded. The target pair is always discarded, as it is
  projected onto a product state by the bilateal measurement.
\end{enumerate}

By a straigtforward calculation, one gets the result that the state
of the remaining pair is still a Bell diagonal state, with the
diagonal coefficients \cite{deutsch96}
\begin{equation}
  \label{eq:qpa_recursion}
  \begin{split}
    A' &= \frac{A^2 + B^2}{N},\quad  B' = \frac{2CD}{N}\\ 
    C' &= \frac{C^2 + D^2}{N},\quad  D' = \frac{2AB}{N},\\
  \end{split}
\end{equation}
and the normalization coefficient \(N= (A+B)^2 + (C+D)^2\), which is
the probability that Alice's and Bob's measurement results in step 3
coincide. Note that, up to the normalization, these recurrence
relations are a quadratic form in the coefficients $A, B, C, $ and
$D$. These relations allow for the following interpretation (which can
be used to obtain the relations (\ref{eq:qpa_recursion}) in the
first place): As all pairs are in the Bell diagonal state
(\ref{eq:bell_diagonal}), one can interpret \(A, B, C,\) and \(D\) as
the relative frequencies in the ensemble of all pairs of the states
\(\Ket{\Phi^+}, \Ket{\Psi^-}, \Ket{\Psi^+},\) and \(\Ket{\Phi^-}\),
respectively. By looking at (\ref{eq:qpa_recursion}) one finds that
the result of combining two $\Ket{\Phi^+}$ or two $\Ket{\Psi^-}$ pairs
is a $\Ket{\Phi^+}$ pair, combining a $\Ket{\Psi^+}$ and a
$\Ket{\Phi^-}$ (or vice versa) yields a $\Ket{\Psi^-}$ pair, and so
on. Combinations of $A, B, C,$ and $D$ that do not occur in
(\ref{eq:qpa_recursion}), namely $AC$, $AD$, $BC$ and $BD$, are
``filtered out'', \ie they give different measurement results for the
bilateral measurement in step 3 of the protocol. We will use this way of
calculating recurrence relations for more complicated situations
later.

Numerical calculations \cite{deutsch96} and, later, an analytical
investigation \cite{macciavello98} have shown that for all initial
states (\ref{eq:bell_diagonal}) with \(A > 1/2\), the recurrence
relations (\ref{eq:qpa_recursion}) approach the fixpoint \(A=1, B = C
= D = 0\); this means that given a sufficiently large number of initial
pairs, Alice and Bob can distill asymptotically pure EPR  pairs.

\section{Factorization of Eve}
\label{sec:fact_of_eve}
In the previous section it has been assumed that Alice and Bob have
perfect apparatus at their disposal, which they use to execute the
protocol. For the following security analysis, we shall consider a
more general scenario where this assumption is abandoned. As mentioned
in the introduction, there is an upper bound \(F_\mathrm{max}\) for
the attainable fidelity of the distilled pairs, when the apparatus
used by Alice and Bob is noisy \cite{briegel,duer_briegel}. For
quantum cryptography the question arises: can these imperfect pairs
still be used for secure key distribution?

In this section we will show that 2--EPP with noisy apparatus is
sufficient to factor out Eve in the Hilbertspace of Alice, Bob, their
labs and Eve. For the proof, we will first introduce the concept of
the lab demon as a simple model of noise. Then we will consider the
special case of binary pairs, where we have obtained analytical
results. Using the same techniques, we generalize the result to the
case of Bell diagonal ensembles. The most general case of correlated
non Bell-diagonal can be reduced to the case of Bell diagonal states
\cite{aschauer2000}.

\subsection{The effect of noise}
\label{sec:eff_of_noise}
In this section we will answer the following question: what is the
effect of an error, introduced by some noisy operation at a given
point of the distillation process? To keep the argument transparent,
we restrict our attention to the following type of noise:
\begin{itemize}
\item It acts locally, \ie noise does not introduce correlations
  between remote quantum systems.
\item Noise is memoryless, \ie on a timescale imposed by the sequence
  of steps in a given protocol, there are no correlations between the
  ``errors'' that occur at different times.
\end{itemize}

The action of noisy apparatus on a quantum system in state \(\rho \in
\rhospace\) can be formally described by some trace conserving,
completely positive map. Any such map can be written in the
operator-sum representation
\cite{kraus_states,schumacher_noisy_channel},
\begin{equation}
  \label{eq:operator_sum}
  \rho \rightarrow \sum_i A_i \rho A_i^\dagger,
\end{equation}
with linear operators \(A_i\), that fulfill the normalization
condition \(\sum_i A A^\dagger = \Eins\). The operators \(A_i\) are
the so-called \emph{Kraus operators} \cite{kraus_states}.

As we have seen above, in the purification protocol the CNOT
operation, which acts on two qubits $a$ and $b$, plays an important
role. For that reason, it is necessary to consider noise which acts on
a two-qubit Hilbert space \(\mathcal{H} = \mathbb{C}^2_a \otimes
\mathbb{C}^2_b\). Eq.~(\ref{eq:operator_sum}) describes the most general
non-selective operation that can, in principle, be implemented. For
technical reasons, however, we restrict our attention to the case that
the Kraus operators are proportional to products of Pauli
matrices. The reason for this choice is that Pauli operators map Bell
states onto Bell states, which will allow us to introduce the very
useful concept of \emph{error flags}
later. Eq.~(\ref{eq:operator_sum}) can then be written as
\begin{equation}
\rho_{ab} \rightarrow  \sum_{\mu,\nu=0}^{3}
f_{\mu\nu}\sigma_{\mu}^{(a)}\sigma_{\nu}^{(b)} \rho_{ab}\sigma_{\mu}^{(a)}
\sigma_{\nu}^{(b)}\,,
\label{eq:noise_model}
\end{equation}
with the normalization condition \(\sum_{\mu,\nu=0}^{3} f_{\mu\nu} =
1\). Note that Eq.~(\ref{eq:noise_model}) includes, for an appropriate
choice of the coefficients $f_{\mu\nu}$, the one- and two-qubit
depolarizing channel and combinations thereof, as studied in
\cite{briegel,duer_briegel}; but it is more general. Below, we will
refer to these special Kraus operators as \emph{error operators}.

The coefficients $f_{\mu\nu}$ can be interpreted as the joint
probability that the Pauli rotations $\sigma_\mu$ and $\sigma_\nu$
occur on qubits $a$ and $b$, respectively. For pedagogic purposes we
employ the following interpretation of (\ref{eq:noise_model}): Imagine
that there is a (ficticious) little demon in Alice's laboratory -- the
``lab demon'' -- which applies in each step of the distillation
process randomly, according to the probability distribution
\(f_{\mu\nu}\), the Pauli rotation \(\sigma_\mu\) and \(\sigma_\nu\)
to the qubits $a$ and $b$, respectively. The lab demon summarizes all
relevant aspects of the lab degrees of freedom involved in the noise
process.

Noise in Bob's laboratory, can, as long as we restrict ourselves to
Bell diagonal ensembles, be attributed to noise introduced by Alice's
lab demon, without loss of generality. It is
also possible to think of a second lab demon in Bob's lab who acts similarly
to Alice's lab demon. This would, however, not affect the arguments
employed in this paper.

The lab demon does not only apply rotations randomly, he also
maintains a list in which he keeps track of which rotation he has
applied to which qubit pair in which step of the distillation
process. What we will show in the following section is that, from the
mere content of this list, the lab demon will be able to extract -- in
the asymptotic limit -- full information about the state of each
residual pair of the ensemble. This will then imply that, given the
lab demons knowledge about the flags, the state of the distilled
ensemble is a tensor product of pure Bell states. Eve cannot not have
information on the specific sequence of Bell pairs (in addition to
their relative frequencies) --- otherwise she would also be able to
learn, to some extent, at which stage the lab demon has applied which
rotation.

From that it follows that Eve is factored out, \ie Alice and Bob
describe their pairs with the state
\begin{equation}
  \label{eq:rho_ABEL}
  \rho_\mathrm{ABEL} = 
      \left( 
        \sum_{i,j = 0}^1 f^{(i,j)}_\mathrm{L}
        \ProjInd{\bell{i,j}}{\mathrm{AB}} 
      \right) 
  \otimes \rho_\mathrm{E}.
\end{equation}

It remains to be shown that the same argument applies to a realistic
scenario where the lab demon is replaced by some ``real'' noise
source. To be specific, in the following argument we show that
\emph{all} quantum or classical devices that share the same noise
characteristics are equally secure. First we note that a communication
protocol is secure if and only if there exists no eavesdropping
strategy; this fact can, in priciple, be determined by cooperating
communication parties. On the other hand, all devices with identical
noise caracteristics are quantum mechanically described by the same
completely positive trace conserving map, so that a initial state
\(\rho_i\) is mapped onto a final state \(\rho_f\), independend from
the physical realization of the map. This means that there is no way
do distinguish the devices by only looking at the input and output
states. For the case of noisy apparatus that is used for entanglement
purification we get thus the following result: a device the implements
(\ref{eq:noise_model}) with a lab demon cannot be distinguished from
any device that introduces noise due to some ``real'' noise source; in
particular, the devices must lead to the same level of security
(regardless whether or not error flags are measured or calculated by
anybody): otherwise they would be distinguishable.

In order to separate conceptual from technical considerations and to
obtain analytical results, we will first concentrate on the special
case of binary pairs and a simplified error model. After that, we
generalize the results to ensembles which are diagonal in the Bell
basis.

\subsection{Binary pairs}
\label{sec:binary_pairs}
In this section we restrict our attention to pairs in the state 
\begin{equation}
  \label{eq:binary_pairs}
  \rho_{AB} = A \ProjInd{\Phi^+}{AB} + B \ProjInd{\Psi^+}{AB},
\end{equation}
and to errors of the form
\label{sec:any_initial_state}
\begin{equation}
  \label{eq:binary_noise}
  \rho_{AB}^{(1)} \otimes \rho_{AB}^{(2)} \rightarrow
  \sum_{\mu,\nu \in \{0,1\}}f_{\mu\nu}U^{(1)}_\mu U^{(2)}_\nu \rho_{AB}^{(1)}
  \otimes \rho_{AB}^{(2)} {U^{(1)}_\mu}^\dagger {U^{(2)}_\nu}^\dagger
\end{equation}
with \(U_0^{(1,2)} = \text{id}^{(1,2)}\) and \(U_1^{(1,2)} =
{\sigma_x}^{(1,2)}\). Eq.~(\ref{eq:binary_noise}) describes a
\emph{two-bit correlated spin-flip channel}. The indices 1 and 2
indicate the source and target bit of the bilateral CNOT (BCNOT)
operation, respectively. It is straightforward to show that, using
this error model in the 2--EPP, binary pairs will be mapped onto
binary pairs.

At the beginning of the distillation process, Alice and Bob share an
ensemble of pairs described by (\ref{eq:binary_pairs}). Let us imagine
that the lab demon attaches one classical bit to each pair, which he
will use for book-keeping purposes. At this stage, all of these bits,
which we call ``error flags'', are set to zero. This reflects the
fact that the lab demon has the same \emph{a priori} knowledge about
the state of the ensemble as Alice and Bob.

In each purification step, two of the pairs are combined. The lab
demon first simulates the noise channel (\ref{eq:binary_noise}) on
each pair of pairs by the process described. Whenever he applies a
\(\sigma_x\) operation to a qubit, he inverts the error flag of the
corresponding pair. Alice and Bob then apply the 2--EPP to each pair
of pairs; if the measurement results in the last step of the protocol
coincide, the source pair will be kept. Obviously, the error flag of
that remaining pair will also depend on the error flag of the the
target pair, \ie the error flag of the remaining pair is a function of
the error flags of both ``parent'' pairs, which we call the \emph{flag
update function}. Since for binary pairs the error flag consists of
one bit, there are 16 different flag update functions in total. From
these, the lab demon chooses the logical AND function as the flag
update function, \ie the error flag of the remaining pair is set to
``1'' if and only if both parent's error flags had the value ``1''. 

After each purification step, the lab demon  divides all pairs into
two subensembles, according to the value of their error flags. By a 
straightforward calculation, one finds for the coefficients $A_i$ and
$B_i$, which completely describe the state of the pairs in the
subensemble $i$, the following recurrence relations:

\begin{equation}
  \label{eq:recursion_binary}
  \begin{split}
    A_0' =  & \frac{1}{N}
    (f_{00}(A_0^2 + 2A_0A_1) + f_{11}(B_1^2+2B_0B_1)\\
    & +f_s(A_0B_1+A_1B_1+A_0B_0))\\
    A_1' =  & \frac{1}{N}
    \left(f_{00}A_1^2 + f_{11}B_0^2 + f_s A_1B_0 \right)\\
    B_0' =   &\frac{1}{N}
    (f_{00}(B_0^2 + 2B_0B_1) + f_{11}(A_1^2+2A_0A_1)\\
    &+f_s(B_0A_1+B_1A_1+B_0A_0))\\
    B_1' =  & \frac{1}{N}
    \left(f_{00}B_1^2 + f_{11}A_0^2 + f_s B_1A_0 \right)
  \end{split}
\end{equation}
with  \(N = (f_{00}+f_{11})((A_0+A_1)^2+(B_0+B_1)^2) + 2f_s (A_0+A_1)
(B_0+B_1) \) and \(f_s = f_{01} + f_{10}\). 

For the case of uncorrelated noise, \(f_{\mu\nu} = f_\mu f_\nu\), we
find an analytical expression for the relevant fixpoint of the map
(\ref{eq:recursion_binary}):
\begin{equation}
  \label{eq:binary_pairs_fixpoint}
  \begin{split}
    A_0^\infty & = \frac
    {4f_0^2 - 4f_0 + (2f_0 - 1)\sqrt{4f_0-3}+1}
    {2 (2 f_0 - 1)^2}, \\
    A_1^\infty & = 0,\quad
    B_0^\infty   = 0 ,\quad
    B_1^\infty   = 1 - A_0^\infty  .  
  \end{split}
\end{equation}

\newcommand{\pd}[2]{\frac{\partial #1}{\partial #2}}

Note that, while Eq. (\ref{eq:binary_pairs_fixpoint}) gives a fixpoint
of (\ref{eq:recursion_binary}) for \(f_0 \ge 3/4\), this does not
imply that this fixpoint is an attractor. In order to investigate the
attractor properties, we calculate the eigenvalues of the matrix of
first derivatives,
\begin{equation}
  \label{eq:derive_binary}
  M_D = \left. \left(
      \begin{array}{c c c}
        \pd{A_0'}{A_0} &\cdots &\pd{B_1'}{A_0}\\ 
        \vdots & \ddots & \vdots \\
        \pd{A_0'}{B_1} &\cdots &\pd{B_1'}{B_1}\\ 
      \end{array}
      \right) \right|_\mathrm{fixpoint} \quad .
\end{equation}

We found that the modulus of the eigenvalues of this matrix is smaller
than unity for \(f_0^\mathrm{crit}=0.77184451 < f_0 \le 1\), which
means that in this interval, the fixpoint
(\ref{eq:binary_pairs_fixpoint}) is also an attractor. This is in
excellent agreement with a numerical evaluation of
(\ref{eq:recursion_binary}), where we found that \(0.77182 <
f_0^\mathrm{crit} < 0.77188\).

\begin{figure}[htbp]
  \begin{center}
    \includegraphics{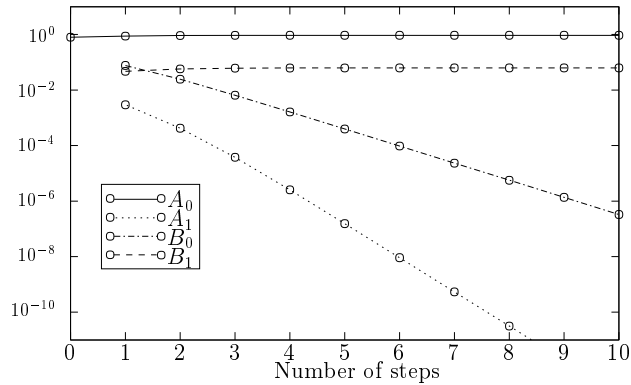}
    \caption{ \label{fig:binary_data}
      The evolution of the four parameters $A_0, A_1, B_0$, and
      $B_1$. Note that both $A_1$ and $B_0$ decrease exponentially
      fast in the number of steps. }
  \end{center}
\end{figure}

We also evaluated (\ref{eq:recursion_binary}) numerically in order to
investigate correlated noise (see Fig.~\ref{fig:binary_data}). Like in
the case of uncorrelated noise, we found that the coefficients $A_0$
and $B_1$ reach, during the distillation process, some finite value,
while the coefficients $A_1$ and $B_0$ decrease exponentially fast,
whenever the noise level is moderate.

In other words, both subensembles, characterized by the value of the
respective error flags, approach a pure state asymptotically. The
pairs in the ensemble with error flag ``0'' are in the state
\(\Ket{\Phi^+}\), while those in the ensemble with error flag ``1''
are in the state \(\Ket{\Psi^+}\).

To conclude this section, we summarize: For a large region
\(0.77184451 \equiv f_0^\mathrm{crit} \le f_0 \le 1\) the 2--EPP
purifies and at the same time any eavesdropper is factored out.  For a
small region \(0.75 < f_0 < f_0^\mathrm{crit}\equiv 0.77184451\),
close to the threshold of the purification protocol, the conditional
fidelity does not reach unity, while the protocol is in the
purification regime. Even though the region is small and of little
practical relevance (in this regime we are already out of the repeater
regime \cite{briegel} and purification is very inefficient), its
existence shows that the process of factorization is an independent
phenomenon and not trivially connected to EPP. A more exhaustive
description of the intermediate regime will be published elsewhere.

\subsection{Bell diagonal initial states}
Now we want to show that the same result is true for arbitrary Bell
diagonal states (Eq.~(\ref{eq:bell_diagonal})) and for noise of the
form (\ref{eq:noise_model}). The procedure is the same as in the case
of binary pairs; however, a few modifications are required.

In order to keep track of the four different error operators
\(\sigma_\mu\) in (\ref{eq:noise_model}), the lab demon has to attach
two classical bits to each pair; let us call them the phase error bit
and amplitude error bit. Whenever a $\sigma_x$ ($\sigma_z$,
$\sigma_y$) error occurs, the lab demon inverts the error amplitude
bit (error phase bit, both error bits). To update these error flags,
he uses the update function given in Tab.~\ref{tab:error_flags}). The
physical reason for the choice of the flag update function will be
given in the next section.

\begin{table}[h]
  \begin{tabular}[t]{|r|cccc|}
    \toprule 
    &(00)&(01)&(10)&(11) \\ 
    \colrule 
    (00)&(00)&(00)&(00)&(10) \\ 
    (01)&(00)&(01)&(11)&(00) \\ 
    (10)&(00)&(11)&(01)&(00) \\
    (11)&(10)&(00)&(00)&(00) \\ 
    \botrule
  \end{tabular}
  \caption
  {The value (phase error,amplitude error) of the updated error flag
  of a pair that is kept after a 2--EPP step, given as a function of the
  error flags of $P_1$ and $P_2$ (left to right and top to bottom,
  respectively). }
  \label{tab:error_flags}
\end{table}

Here, the lab demon divides all pairs into four subensembles,
according to the value of their error flag. In each of the
subensembles the pairs are described by a Bell diagonal density
operator, like in Eq.~(\ref{eq:bell_diagonal}), which now depends on
the subensemble.  That means, in order to completely specify the state
of all four subensembles, there are 16 real numbers
\(A_{ij},B_{ij},C_{ij}, D_{ij}\) with \(i,j \in\{0,1\}\) required, for
which one obtaines  recurrence relations of the form
\begin{equation}
  \label{eq:recurrence}
  \begin{split}
    A^{(00)}_{n} &\rightarrow
    A^{(00)}_{n+1}(A^{(00)}_n,A^{(01)}_n,\ldots,D^{(11)}_n),\\
    A^{(01)}_{n} &\rightarrow
    A^{(01)}_{n+1}(A^{(00)}_n,A^{(01)}_n,\ldots,D^{(11)}_n),\\
    &\,\,\,\vdots \\ D^{(11)}_{n} &\rightarrow
    D^{(11)}_{n+1}(A^{(00)}_n,A^{(01)}_n,\ldots,D^{(11)}_n).\\
  \end{split}
\end{equation}
These generalize the recurrence relations (\ref{eq:recursion_binary})
for the case of binary pairs, and the relations
(\ref{eq:qpa_recursion}) for the case of noiseless apparatus.

Like the recurrence relations (\ref{eq:qpa_recursion}) and
(\ref{eq:recursion_binary}), respectively, these relations are (modulo
normalization) quadratic forms in the 16 state variables \(\vec {a} =
\left(A^{(00)}, A^{(00)}, \ldots, D^{(11)} \right)^T\), with
coefficients that depend on the error parameters \(f_{\mu\nu}\)
only. In other words, (\ref{eq:recurrence}) can be written in the more
compact form
\begin{equation}
  \label{eq:quadratic_form}
  \vec{a}'_j = \vec{a} M_j \vec{a}^T,
\end{equation}
where, for each \(j \in \{1,\ldots 16\}\), \(M_j\)  is a real \(16\times
16\)-matrix whose coefficients are polynomials in the noise parameters
\(f_{\mu\nu}\).

\subsection{Numerical results}
\label{sec:results_full}
The 16 recurrence relations (\ref{eq:recurrence}) imply a reduced set
of 4 recurrence relations for the quantities $A_n=\sum_{ij}A_n^{(ij)},
\ldots, D_n=\sum_{ij}D_n^{(ij)}$ which describe the evolution of the
total ensemble (that is, the \emph{blend} \cite{englert1999} of the
four subensembles) under the purification protocol. Note that these
values are the only ones which are known and accessible to Alice and
Bob, as they have no knowledge of the values of the error flags. It
has been shown in \cite{briegel} that under the action of the noisy
entanglement distillation process, these quantities converge towards a
fixpoint $(A_\infty, B_\infty, C_\infty, D_\infty)$, where
$A_\infty=F_{\text{max}}$ is the maximal attainable fidelity
\cite{duer_briegel}.

\begin{figure}[tp]
  \begin{center}
    \includegraphics[width=.3\figurewidth]{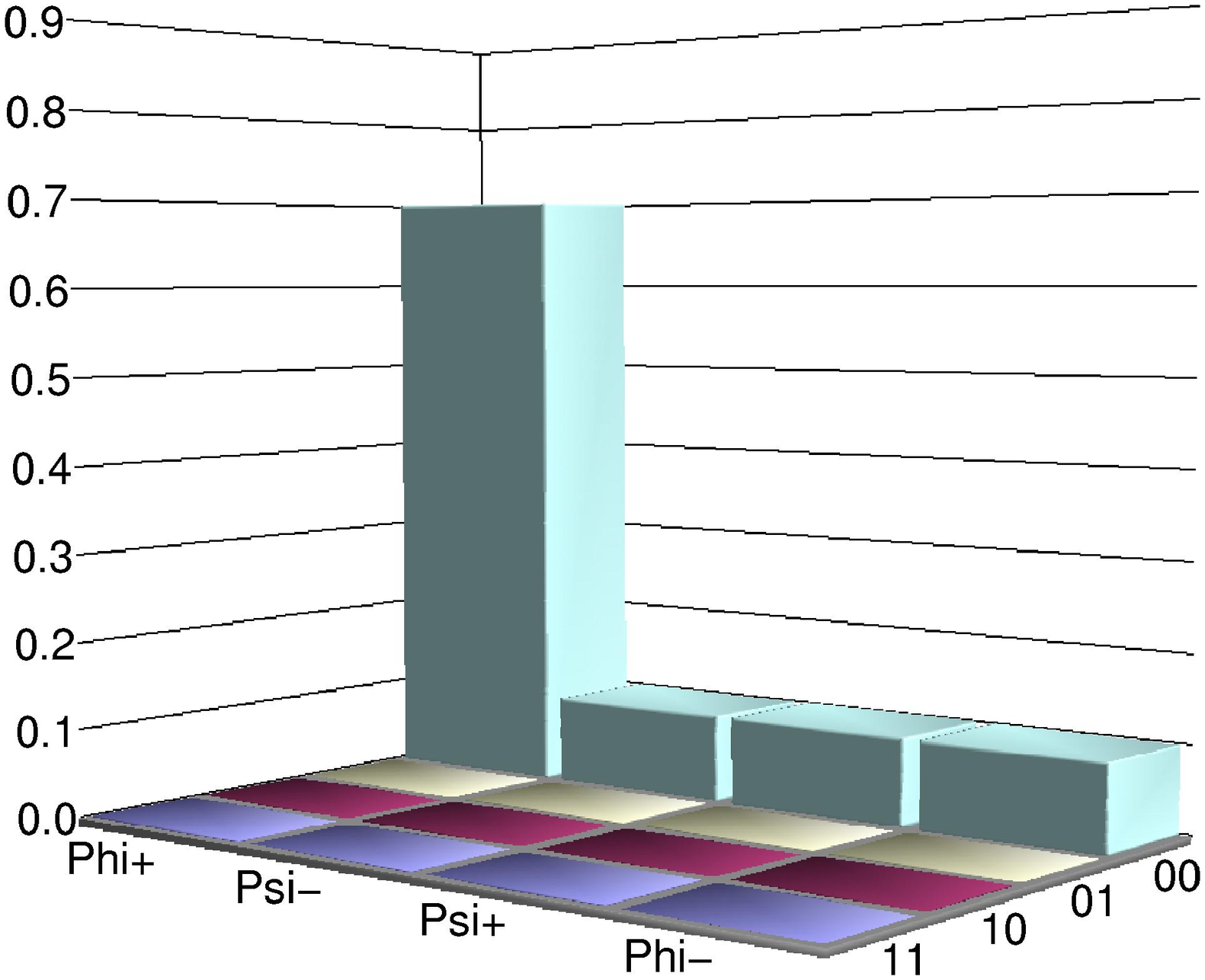}%
    \includegraphics[width=.3\figurewidth]{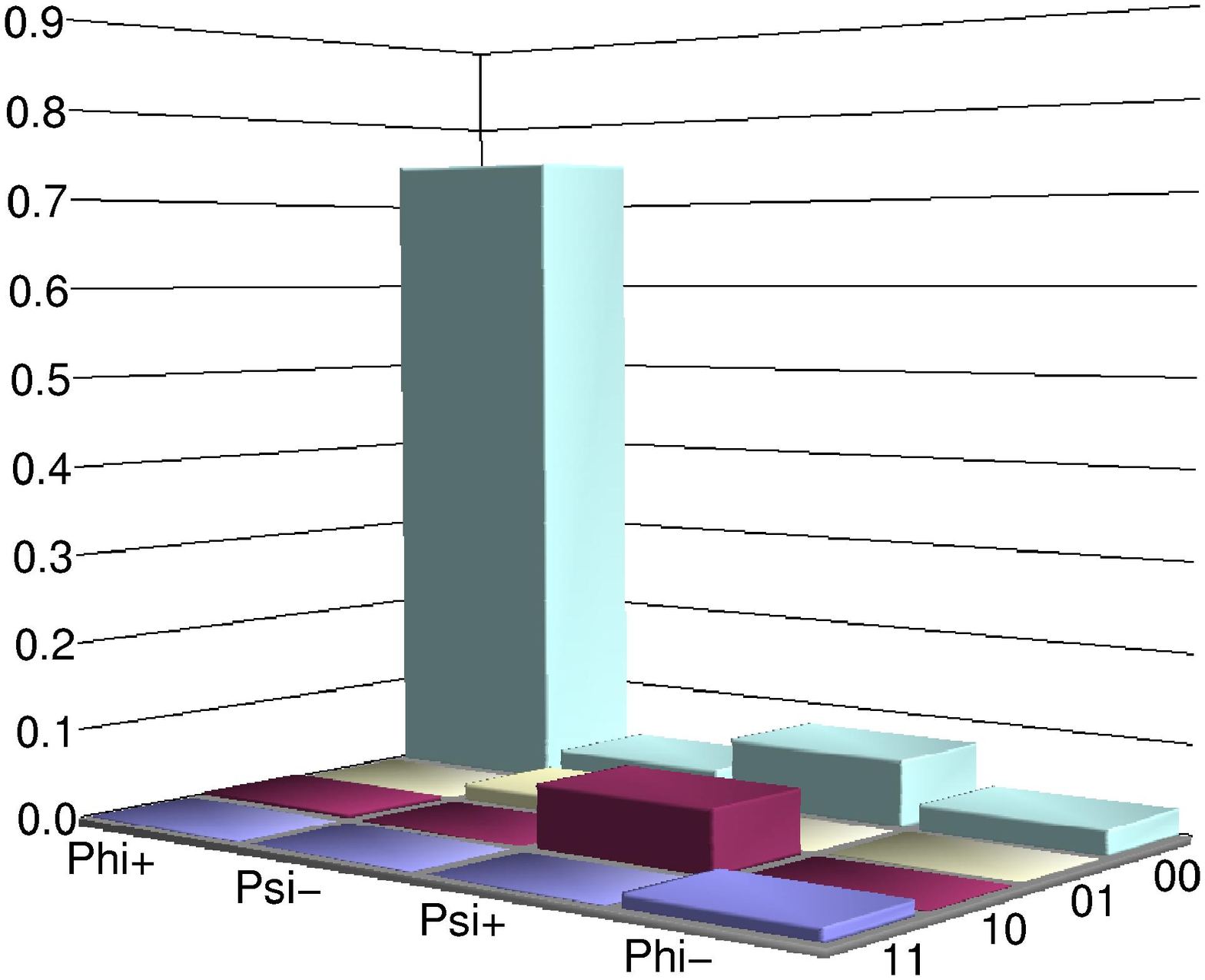}%
    \includegraphics[width=.3\figurewidth]{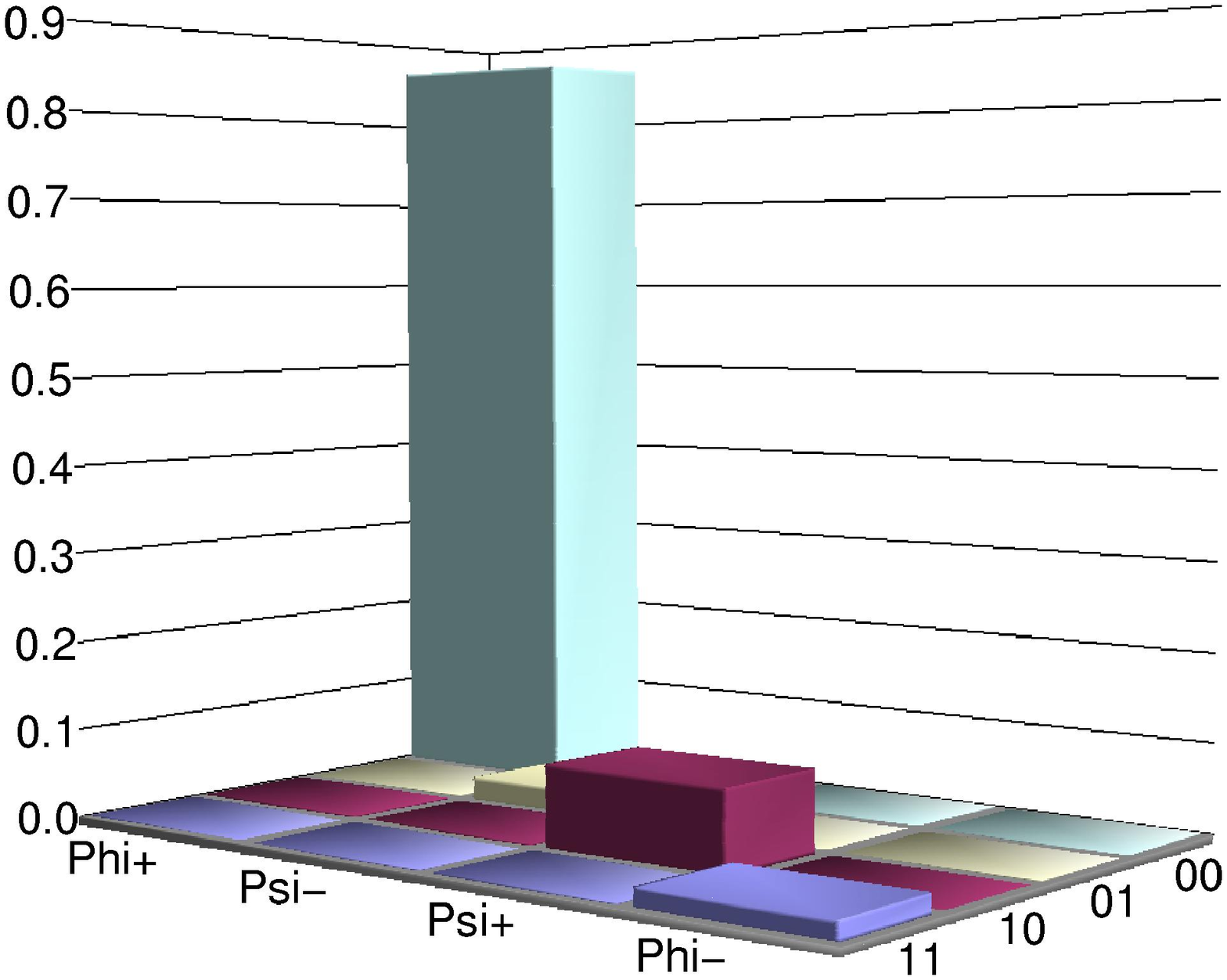}
    \caption{    \label{fig:state_evolution}
      The evolution of the extended state}
  \end{center}
\end{figure}

Fig. \ref{fig:state_evolution} shows for typical initial conditions
the evolution of the 16 coefficients \( A_n^{(00)}\ldots
D_n^{(11)}\). They are organized in a $4\times 4$-matrix, where one
direction represents the Bell state of the pair, and the other
indicates the value of the error flag. The figure shows the state (a)
at the beginning of the entanglement purification procedure, (b) after
few purification steps, and (c) at the fixpoint. As one can see,
initially all error flags are set to zero and the pairs are in a
Werner state with a fidelity of \(70\%\). After a few steps, the
population of the diagonal elements starts to grow; however, none of
the elements vanishes. At the fixpoint, all off-diagonal elements
vanish, which means that there are \emph{strict correlations} between
the states of the pairs and their error flags. 

In order to determine how fast the state converges, we investigate two
important quantities: the first is the fidelity $F_n \equiv A_n$, and
the second is the \emph{conditional fidelity}
\(F_n^{\text{cond}}\equiv A_n^{(00)}+ B_n^{(11)}+ C_n^{(01)}+
D_n^{(10)}\).  Note that the first quantity is the sum over the four
\(\Ket{\Phi^+}\) components in Fig. \ref{fig:state_evolution}, while
the latter is the sum over the four diagonal elements. The conditional
fidelity is the fidelity which Alice and Bob would assign to the pairs
if they knew the values of the error flags, \ie
\begin{equation}
  \label{eq:cond_fidelity}
  F_n^{\text{cond}} = \sum_{i,j} \Bra{\Phi^+} \sigma_{i,j}\rho_{i,j}
  \sigma_{i,j} \Ket{\Phi^+},
\end{equation}
where $\rho_{i,j}$ is the non-normalized state of the subensemble of
the pairs with the error flag $(i,j)$. For convenience, we used the
phase- and spin-flip bits $i$ and $j$ as indices for the Pauli
matrices, \ie \(\sigma_{00} = \mathrm{Id}, \sigma_{01} = \sigma_x,
\sigma_{11} = \sigma_y, \sigma_{10} = \sigma_z\).

\begin{figure}[tp]
  \begin{center}
    \includegraphics[width=0.45\figurewidth]{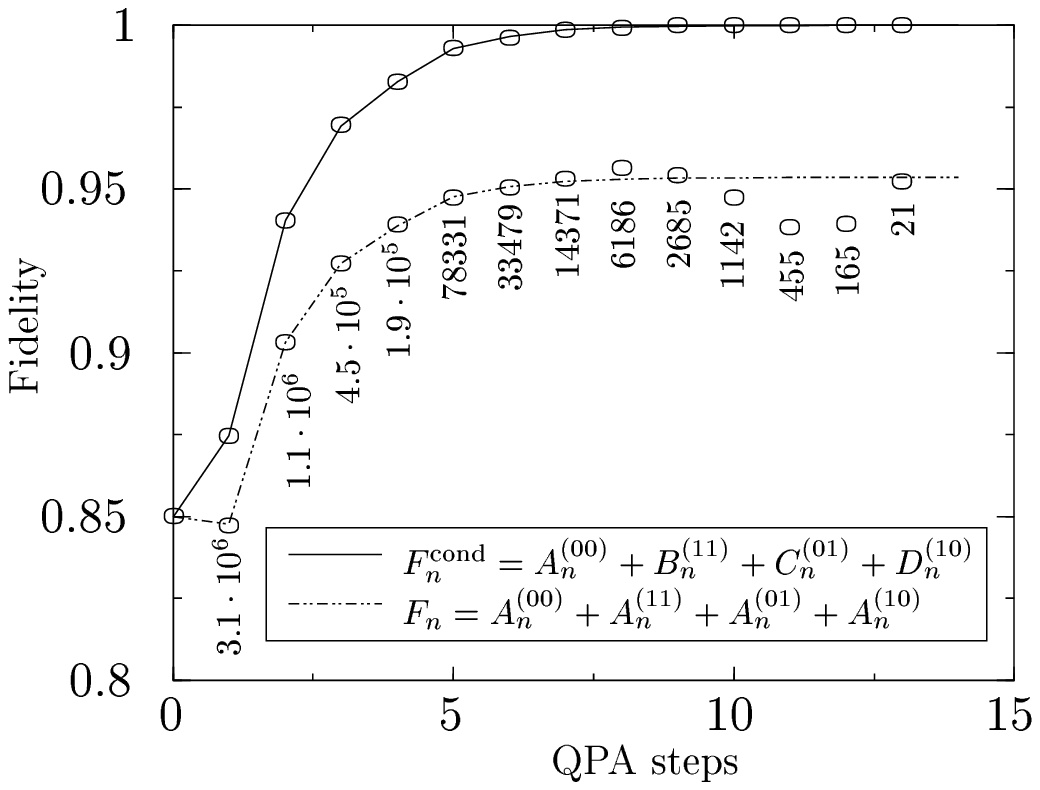}
    \includegraphics[width=0.45\figurewidth]{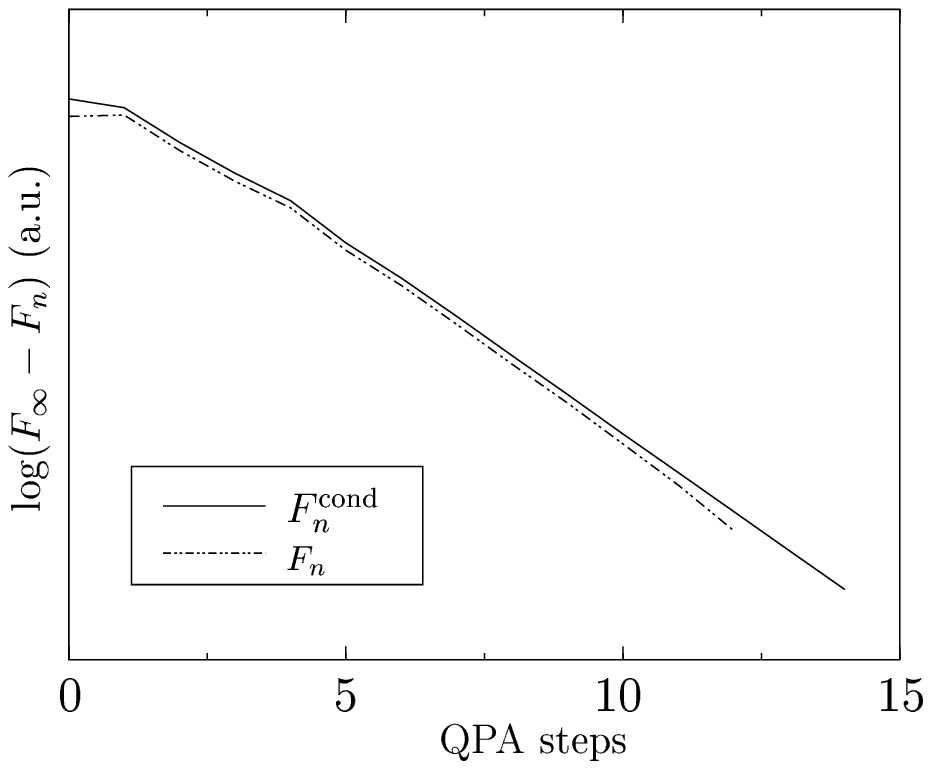}\\
    (a)\hspace{0.45\figurewidth}(b)
    \caption{  \label{fig:F_and_F_cond_N}
      (a) The fidelities $F$ and $F_{\text{cond}}$ as a function of
      the number of steps in the entanglement distillation process
      (analytical results (lines) and Monte Carlo simulation
      (circles)). For the calculation, one and two qubit white noise
      with a noise fidelity of 97\% has been assumed. The Monte Carlo
      simulation was started with 10000000 pairs; the numbers indicate
      how many pairs are left after each step of the distillation
      process. This decreasing number is the reason for the
      increasing fluctuations around the analytical curves.}
  \end{center}
\end{figure}

\begin{figure}[htp]
  \begin{center}
    \includegraphics{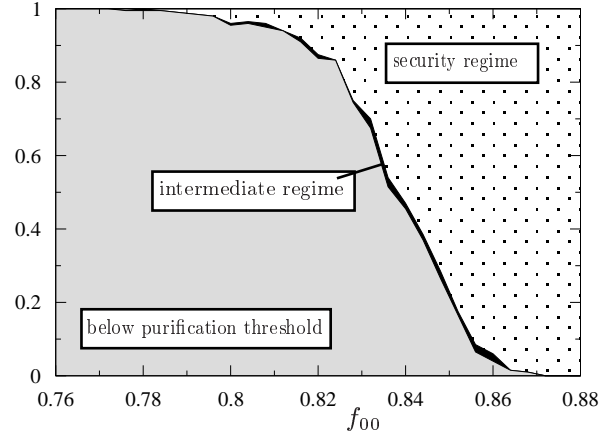}
    \caption{    \label{fig:distri_regimes}
      The size and the location of the three regimes of the
      distillation process. For fixed values of \(f_{00}\), the
      remaining 15 noise parameters \(f_{\mu\nu}\) have been choosen
      at random. Plotted is the relative frequency of finding the
      noise parameters in any of the three regimes as a function of
      \(f_{00}\). } 
  \end{center}
\end{figure}

The results that we obtain are similar to those for the binary
pairs. We can also distinguish three regimes of noise parameters
\(f_{\mu\nu}\). In the high-noise regime (i.\,e., small values of
\(f_{00}\)), the noise level is above the threshold of the 2--EPP and
the fidelity \(F\) and the conditional fidelity \(F^\mathrm{cond}\)
converge both to the value 0.25. In the low-noise regime (i.\,e.,
large values of \(f_{00}\)), F converges to the maximum fidelity
\(F_\mathrm{max}\) \emph{and} \(F^\mathrm{cond}\) converges to unity
(see Fig. \ref{fig:F_and_F_cond_N}). This regime is the \emph{security
regime}, where we know that secure quantum communication is
possible. Like for binary pairs, there exists also an intermediate
regime, where the 2--EPP purifies but \(F^\mathrm{cond}\) does not
converge to unity.  For an illustration, see
Fig~\ref{fig:distri_regimes}. Note that the size of the intermediate
regime is very small, compared to the security regime. Whether or not
secure quantum communication is possible in this regime is
unknown. However, the answer of this question is irrelevant for all
practical purposes, because in the intermediate regime the
distillation process converges very slowly. A detailed discussion of
the situation near the purification threshold will be published at
some other place.

Even though the intermediate regime is practically irrelevant, it is
important to estimate its size. For simplicity, we considered the case
of one-qubit white noise, i.e. \(f_{\mu\nu} = f_\mu f_\nu \) and \(f_1
= f_2 = f_3 = (1-f_0)/3\). Here, this regime is known to be bounded by
\[0.8983 < f^\mathrm{crit,lower} < f_0 < f^\mathrm{crit,upper} <
0.8988.\] 

\begin{figure}[tp]
  \begin{center}
    \includegraphics[width=\figurewidth]{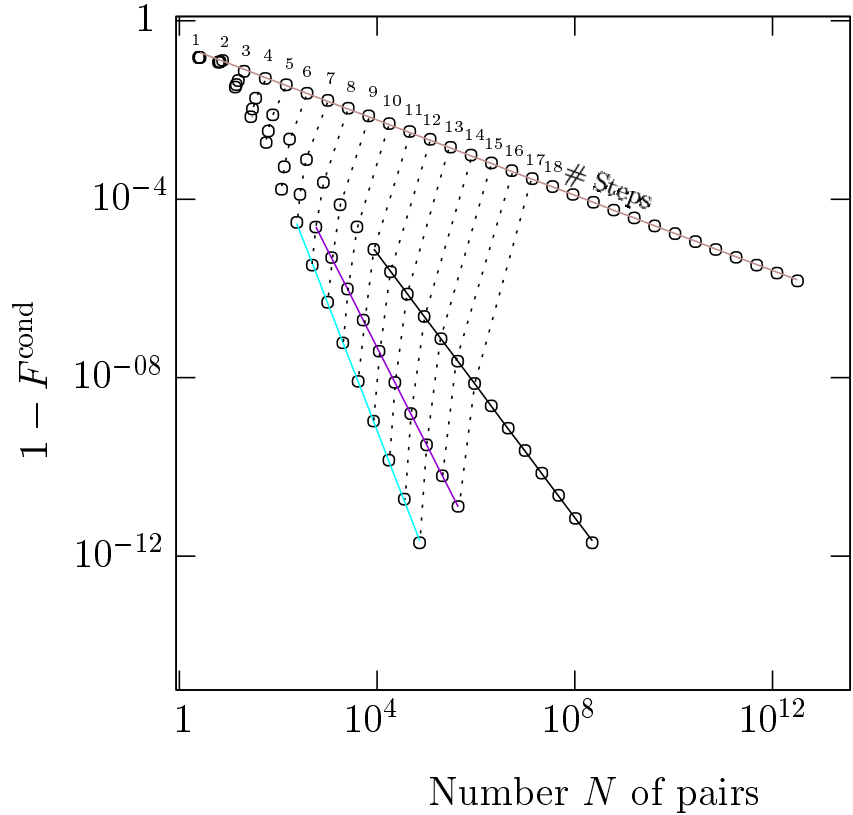}
    \caption{   
      Number $N$ of pairs needed to create one pair with conditional
      fidelity $F^{\rm cond}$. The initial state of the pairs was of
      the Werner type with fidelity $F_0$ = 85\%. One- and two-qubit
      errors were assumed with noise parameters $f_{1\rm bit} =
      f_{2\rm bit} = 0.95, 0.98, 0.99, 0.995$ (from top to bottom).}
    \label{fig:F-N}
  \end{center}
\end{figure}

Regarding the efficiency of the distillation process, it is an
important question how many initial pairs are needed to create one
pair with the \emph{security parameter} \(\epsilon \equiv 1 -
F^{\text{cond}}\). Both the number of required initial pairs
(resources) and the security parameter scale exponentially with the
number of distillation steps, so that we expect a polynomial relation
between the resources and the security parameter
\(\epsilon\). Fig.~\ref{fig:F-N} shows this relation in a log-log plot
for different noise parameters. The straight lines are fitted
polynomial relations; the fit region is indicated by the lines
themselves.

\section{Discussion}
\label{sec:dicussion}
We have shown in the preceeding section, that the two-way entanglement
distillation process is able to disentangle any eavesdropper from Bell
pairs distributed between Alice and Bob, even in the presence of
noise, \ie when the pairs can only be purified up to a specific
maximun fidelity \(F_\mathrm{max} < 1\). Alice and Bob may use them as
a secure quantum communication channel. They are thus able to perform
secure quantum communication, and, as a special case, secure classical
communication (which is in this case equivalent to a key distribution
scheme).

In order to keep the argument transparent, we first considered the
case where noise of the form (\ref{eq:noise_model}) is explicitly
introduced by a fictious lab-demon, who keeps track of all error
operations and performs calculations. However, using a simple
indistinguishability argument (see Section \ref{sec:eff_of_noise}), we
could show that any apparatus with the noise characteristics
(\ref{eq:noise_model}) is equally secure as a situation where noise in
introduced by the lab demon. This means that the security of the
protocol does not depend on the fact whether or not anybody actually
calculates the flag update function. It is sufficient to just use a noisy
2--EPP, in order to get a secure quantum channel.

For the proof, we had to make several assumptions on the noise that
acts in Alices and Bobs entanglement purification device. One
restriction is that we only considered noise which is of the form
(\ref{eq:noise_model}). However, this restriction is only due to
technical reasons; we conjecture that our results are also true for
most general noise models of the form (\ref{eq:operator_sum}). We have
also implicitly introduced the assumption that the eavesdropper has no
additional knowledge about the noise process, \ie Eve only knows the
publicly known noise characteristics (\ref{eq:noise_model}) of the
apparatus. This assumption would not be justified, for example, if the
lab demon was bribed by Eve, or if Eve was able to manipulate the
apparatus in Alice's and Bob's laboratories, for example by shining in
light from an optical fiber. This concern is not important from a
principial point of view, as the laboratories of Alice and Bob are
considered secure by assumption. On the other hand, this concern has
to be taken into account in any practical implementation.

\section*{Acknowledgments}
We thank C.~H.~Bennett, A.~Ekert, G.~Giedke, N.~L\"utken\-haus,
J.~M\"uller-Quade, R.~Rau{\ss}endorf, A.~Schenzle, Ch.~Simon and
H.~Weinfurter for valuable discussions. This work has been supported
by the Deutsche Forschungsgemeinschaft through the
Schwer\-punkts\-programm ``Quanten\-in\-for\-mations\-ver\-ar\-bei\-tung''.



\end{document}